\input{aipcheck}

\documentclass[
    ,final            
  ]
  {aipproc}

\layoutstyle{6x9}

\begin{document}

\title{ The true nature of Terzan 5: the most efficient "furnace" of MSPs in the Galaxy}


\author{Francesco R. Ferraro}{
  address={ Dipartimento di Astronomia, Universit\`a di Bologna Via
  Ranzani 1 -- 40127 Bologna (ITALY)\\
  email: francesco.ferraro3@unibo.it}
}

\begin{abstract}
 Terzan\,5  is the globular cluster (GC)-like stellar system 
  harboring the largest known population of MSPs.
  Using the Multi-Conjugate Adaptive Optics demonstrator MAD
  at the ESO-VLT, we recently obtained 
  a superb ($K,\,J-K$)
  color-magnitude diagram, which has revealed the existence of two horizontal
  branches (HBs) well separated in magnitude and colour
  (Ferraro et al. 2009, Nature,462, 483).  A prompt
  spectroscopic follow-up with NIRSPEC@Keck has shown that the two
  populations have (1) significantly different iron content ([Fe/H]$=-0.2$
  and $+0.3$ for the faint and the bright HB, respectively),  
  (2) distinct  [$\alpha$/Fe] abundance patterns and (3) no
evidence of the Al-O anti-correlation commonly observed in GCs.  
All these  properties suggest that  Ter\,5 is far from being a 
genuine globular. Instead  it has experienced 
the explosion of a huge number of supernovae (SNe), thus accounting for its high metal content 
and it should have been much more massive 
in the past than today, thus to retain
the SN ejecta within its potential well. The many type II SNe
should have also produced a large number of neutron stars (NSs), which could 
finally explain its exceptionally large population of MSPs.  
 \end{abstract}

\maketitle

Terzan\,5 (hereafter Ter\,5)  was discovered in 1968  
and, since then, it was cataloged as a common Globular Cluster (GC) in the Galactic Bulge. 
While it was early
indicated as the GC with the highest collision rate in the Galaxy
\citep{ver87}, the 
astonishing discovery of 31
millisecond pulsars (MSPs) \citep{ransom}
has recently renewed the interest toward this stellar system. 
The updated list (see
http://www.naic.edu/$\sim$pfreire/GCpsr.html) now counts 34 such
objects, corresponding to 25\% of the entire sample of known MSPs in
GCs; indeed \emph{the MSP population of Ter\,5 is the largest ever found in
 any GCs!}

Within a project aimed at studying the properties of GCs harboring MSPs, a 
 few years ago our group started  
a  systematic search of optical companions to MSP in
binary systems \citep{fe01a}. This  search has led to
important results, since 4 (out of the 7) known MSP companions 
in  GCs (see Fig. 1) have been discovered  by our group.
Within this search,
Ter\,5  was, of course, one of the top-priority target,   and 
in fact it was the subject of a deep  photometric investigation.  
In particular, our latest survey
led to a surprising discovery. 
Despite the severe extinction ($\rm E(B-V)=2.4$; \citep{val07}) affecting this system,
thanks to near-IR observations performed with the Multi-conjugate
Adaptive Optics Demonstrator (MAD) at the ESO-VLT, we have obtained a
superb ($\rm K,\,J-K$) color-magnitude diagram (CMD) even for the very
central region of the cluster. This allowed us to reveal the presence
of {\it two well-defined red horizontal branch (HB) clumps, clearly
separated in luminosity} \citep{fe09}: a bright HB (bHB) at $K=12.85$, and a
faint HB (fHB) at $K=13.15$, the latter having a bluer color (see
Fig.2a). Since theoretical models \citep{pie04}  
predict that the HB level gets brighter in
the K-band for increasing metallicity, a combination of 
different metallicities (with the bHB
population being more metal-rich than the fHB) could in principle
reproduce the observed feature.  Prompt medium-resolution, near-IR
spectra (acquired with NIRSPEC at the Keck Telescope) have indeed
confirmed that the iron content of the stars in the two clumps differs
by a factor of 3 ($\sim 0.5$ dex): the fHB stars have [Fe/H]$=-0.2$,
while the bHB stars have [Fe/H]$ =+0.3$ (Fig.2b). Note that
Ter\,5 is the first GC-like system ever discovered in the Galactic bulge
with a clear iron spread\footnote{The only other known globular-like stellar system
with a significant spread in iron abundance (in a much lower metallicity regime)
and multiple stellar populations is $\omega$ Centauri in the galactic Halo
(see \citep{nor,fer04,sol05})}.  However, such a different iron content
alone, cannot account for the observed split in luminosity of the two
HB clumps.  Instead, two populations characterized by the measured
iron abundances and two different ages ($t= 12$ Gyr for the fHB, and
$t=6$ Gyr for the bHB) well reproduce both the luminosity of the two
HB clumps and the location of the red giant branches (RGBs). 
Note that a much smaller age gap is needed if
the younger population is enhanced in Helium \citep{dan10}.
  

\begin{figure}
\includegraphics[height=.3\textheight]{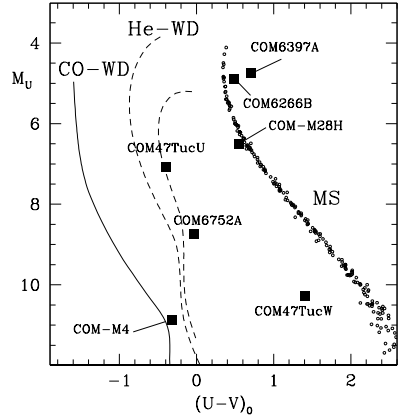}
  \caption{ Location in the ($M_U,\,(U-V)_0$) plane
  of the optical companions ({\it filled squares}) to binary MSPs
  detected so far in GCs: COM47TucU and COM47TucW in 47\,Tucanae
  (\citep{ed01,ed02});
  COM6397A in NGC6397 \citep{fe01b}; COM6752A
  in NGC6752 \citep{fe03}; COM-M4 in M4 \citep{sig03}, COM6266B in NGC6266
  \citep{coco} and COM$-$M28H recently
  discovered in M28 by \citep{pal}.    The  
    He-WDs ({\it dashed lines}) and CO-WD ({\it solid lines}) cooling tracks 
  and Main Sequence stars
  ({\it small empty circles}) are plotted for reference.}
\end{figure}
 

\begin{figure}
\includegraphics[height=.27\textheight]{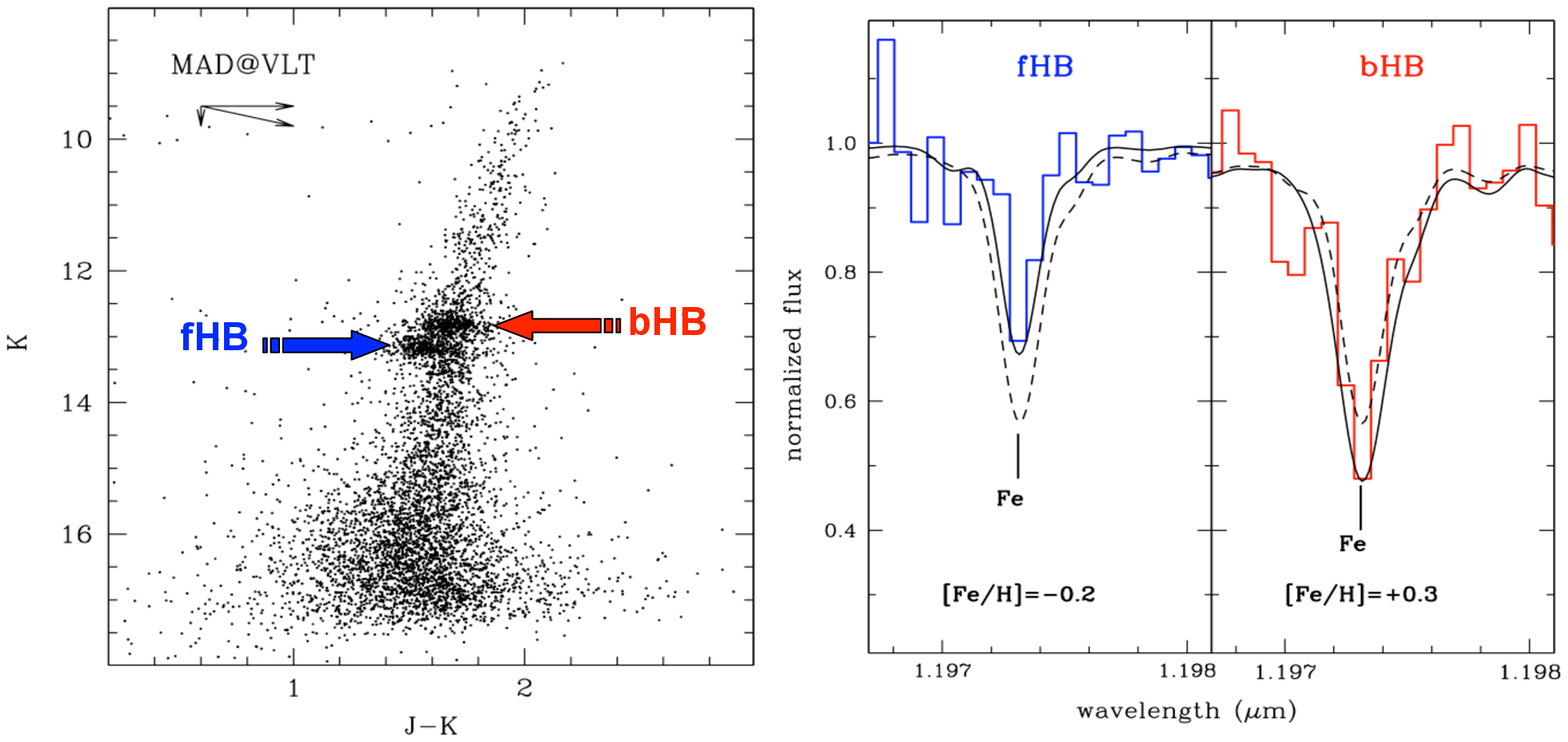}
  \caption{ {\it Left panel --} VLT-MAD (K,J-K) CMD of the
  central ($1' \times 1'$) region of Ter\,5, with the two HB clumps marked with red
  arrows. The reddening vector is also plotted. 
  {\it Right panel --} Combined spectra near the
  1.1973 micron iron line for three fHB (left) and three bHB (right)
  stars, as obtained with NIRSPEC at Keck II. The
  measured equivalent widths of the lines and suitable spectral
  synthesis yield [Fe/H]$\simeq -0.2\pm 0.1$ and
  [Fe/H]$\simeq +0.3 \pm 0.1$ iron abundances, respectively.}
\end{figure}

Additional high-resolution spectra of a sample of RGB
stars \citep{ori10} confirmed the striking difference 
in the iron content of the two populations. Moreover, we found 
indications of distinct [$\alpha$/Fe] abundance patterns and no
evidence of the Al-O anti-correlation commonly observed in GCs.  These
chemical properties suggest that Ter\,5 is not a genuine
GC. Indeed, the
co-existence of two stellar populations with different iron
content (and probably ages) indicates that the original mass of
Ter\,5 was significantly larger in the past than observed today,
large enough to retain the iron-enriched gas that, otherwise, would
have been ejected out from the system by the violent supernova (SN)
explosions.  Indeed, the smallest systems with solid evidences of a
spread in the iron content (and ages) are significantly more massive
than GCs: the dwarf spheroidal satellites of the Milky Way typically
have masses of $\sim 10^7 M_\odot$  \citep{stri08}
with initial masses possibly amounting to a
few $10^8 M_\odot$ \citep{rev09}. 

The exceptionally high metallicity regime of the two stellar
populations found in Ter\,5 also suggests a quite efficient
enrichment process, that could have a relevant role in the origin of
its population of MSPs. In particular, both the iron and the
[$\alpha/$Fe] abundance ratios measured in Ter\,5  \citep{ori10} 
show a remarkable similarity
with those of the Bulge stars.  {\it This strongly suggests that
  Ter\,5 and the Galactic Bulge shared the same star formation and
  chemical enrichment processes.}  The many observations of Bulge
stars \citep{ful07,zoc08}  indicate that they are all
characterized by an old age, a high (close to solar) average
metallicity [Fe/H], and an [$\alpha/$Fe] ratio which is enhanced (due
to SNII enrichment)\footnote{The [$\alpha/$Fe]--[Fe/H] relation shows
  a down-turn at a value of [Fe/H] which depends on the star formation
  rate: the higher the latter, the higher the metallicity at which the
  down-turn occurs. Such a value is [Fe/H]$\simeq -1$ in the Old
  Halo/Disk, while it is significantly higher ([Fe/H]$\simeq 0$) in
  the Bulge, testifying a much higher star formation rate in this
  dense environment.}  up to a metallicity [Fe/H]$\simeq 0$. These
constraints suggest a scenario where the dominant stellar population
of the Bulge formed early (thus explaining the old age), rapidly and
with high efficiency from a gas mainly enriched by SNII (thus
explaining the [$\alpha/$Fe] enhancement up to high iron contents).
Also chemical evolution models \citep{bal} 
indicate that the abundance patterns observed in the Bulge require a
quite high star formation efficiency and an initial mass function
flatter than that in the solar neighbourhood, thus to rapidly enrich
the gas up to about solar metallicity through an exceptionally large
amount of SNII explosions.

The assumption of a similar scenario for Ter\,5 would naturally
explain its extraordinary population of MSPs, since the expected high
number of SNII would produce a large population of neutron stars (NSs), most
of which would have been retained by the deep potential well of the
massive proto-Ter\,5 system.  
 Then, the large collision rate of Ter\,5 \citep{ver87,lan10} could have favored the formation of
binary systems containing NSs and promoted the re-cycling process that
finally generated the large population of MSPs we observe today.  
  If such a scenario is
correct, many more MSPs still wait to be discovered in this system, 
and the 34 known objects probably are just
the tip of the iceberg. 
This is also supported by recent   
high-energy observations  with the Fermi Space Telescope that 
discovered GeV $\gamma$-ray emission  from Ter\,5  \citep{kon}
and 47 Tuc \citep{abdo}.  
The  $\gamma$-ray luminosity attributed to MSPs indicates that Ter\,5 may
 host a factor of 5-20 more MSPs (depending on the actual
cluster distance) than 47~Tuc (counting 23 such objects).
Future deeper pulsar searches of Ter\,5,
perhaps with larger telescopes such as the Square Kilometer Array,
will shed additional light on the nature of this system.


\begin{theacknowledgments}
This research was supported by  ASI (under
contract ASI-INAF I/009/10/0), by INAF (under contract PRIN-INAF2008) and by the 
MIUR.

\end{theacknowledgments}



\bibliographystyle{aipprocl} 





\end{document}